# Mechanical Characterization of Brain Tissue in Tension at Dynamic Strain Rates

Badar Rashid[a], Michel Destrade[b,a], Michael D Gilchrist[a*]

[a]School of Mechanical and Materials Engineering, University College Dublin, Belfield, Dublin 4, Ireland

[b]School of Mathematics, Statistics and Applied Mathematics, National University of Ireland Galway, Galway, Ireland

*Corresponding Author

Tel: + 353 1 716 1884/1991, + 353 91 49 2344  Fax: + 353 1 283 0534

Email: Badar.Rashid@ucdconnect.ie (B. Rashid), michael.gilchrist@ucd.ie (M.D. Gilchrist), michel.destrade@nuigalway.ie (M. Destrade)

**Abstract**       Mechanical characterization of brain tissue at high loading velocities is crucial for modelling Traumatic Brain Injury (TBI). During severe impact conditions, brain tissue experiences compression, tension and shear. Limited experimental data is available for brain tissue in extension at dynamic strain rates. In this research, a High Rate Tension Device (HRTD) was developed to obtain dynamic properties of brain tissue in extension at strain rates of ≤ 90/s. *In vitro* tensile tests were performed to obtain properties of brain tissue at strain rates of 30, 60 and 90/s up to 30% strain. The brain tissue showed a stiffer response with increasing strain rates, showing that hyperelastic models are not adequate. Specifically, the tensile engineering stress at 30% strain was 3.1±0.49 kPa, 4.3±0.86 kPa, 6.5±0.76 kPa (mean ± SD) at strain rates of 30, 60 and 90/s, respectively. Force relaxation tests in tension were also conducted at different strain magnitudes (10% - 60% strain) with the average rise time of 24 ms, which were used to derive time dependent parameters. One term Ogden, Fung and Gent models were used to obtain material parameters from the experimental data. Numerical simulations were performed using a one-term Ogden model to analyze hyperelastic behavior of brain tissue up to 30% strain. The material parameters obtained in this study will help to develop biofidelic human brain finite element models, which can subsequently be used to predict brain injuries under impact conditions and as a reconstruction and simulation tool for forensic investigations.

# 1 Introduction

Traumatic brain injury (TBI) occurs when a sudden trauma causes damage to the brain. The damage can be focal, confined to one area of the brain, or diffuse, involving more than one area of the brain. Symptoms of a TBI can be mild, moderate, or severe, depending on the extent of the damage to the brain. Most TBIs are due to road transportation accidents but they also occur from workplace or sports accidents and from assaults (O'Riordain et al., 2003; Forero Rueda & Gilchrist, 2009). *Concussion* is the most minor and the most common type of TBI, whereas *diffuse axonal injury* (DAI) is the most severe form of injury which involves damage to individual nerve cells (*neurons*) and loss of connections among neurons. To gain a better understanding of the mechanisms of TBI, several research groups have developed numerical models which contain detailed geometric descriptions of anatomical features of the human head, in order to investigate internal dynamic responses to multiple loading conditions (Ho and Kleiven, 2009; Horgan and Gilchrist, 2003; Kleiven, 2007; Kleiven and Hardy, 2002; Raul et al., 2006; Ruan et al., 1994; Takhounts et al., 2003; Zhang et al., 2001). However, the biofidelity of these models is highly dependent on the accuracy of the material properties used to model biological tissues, therefore more systematic research on the constitutive behavior of brain tissue under impact is essential.

Over the past three decades, several research groups investigated the mechanical properties of brain tissue in order to establish constitutive relationships over a wide range of loading conditions. Mostly dynamic oscillatory shear tests were conducted (Arbogast et al., 1997; Bilston et al., 2001; Brands et al., 2004; Darvish and Crandall, 2001; Fallenstein et al., 1969; Hrapko et al., 2006; Nicolle et al., 2004; Nicolle et al., 2005; Prange and Margulies, 2002; Shuck and Advani, 1972; Thibault and Margulies, 1998) and unconfined compression tests (Cheng and Bilston, 2007; Estes and McElhaney, 1970; Gilchrist, 2004; Miller and Chinzei, 1997; Pervin and Chen, 2009; Prange and Margulies, 2002; Rashid et al., 2012b; Tamura et al., 2007). However, only a limited number of tensile tests has been conducted (Miller and Chinzei, 2002; Tamura et al., 2008; Velardi et al., 2006). To the authors' knowledge, no experimental data for brain tissue in tension at dynamic strain rates is available except for that of Tamura et al., (2008), who performed tests at 0.9, 4.3 and 25/s, where the fastest rate was closest to impact speeds.

Considering the difficulty of obtaining human brain tissue for *in vitro* testing, experiments are usually performed on animal brain samples (monkey, porcine, bovine, rabbit, calf, rat or mouse). Galford and McElhaney (1970) showed that shear, storage and loss moduli are 1.5, 1.4 and 2 times higher, respectively for monkeys than for humans. Similarly, Estes and McElhaney (1970) performed tests on human and Rhesus monkey tissue and found that the response of the Rhesus monkey tissue was slightly higher than the response of human brain tissue at comparable compression rates. Differences between human and porcine brain properties were also pointed out by Prange et al., (2000), who demonstrated that human brain tissue stiffness was 1.3 times higher than that of porcine brain. However, Nicolle et al., (2004) observed no significant difference between the mechanical properties of human and porcine brain matter. Pervin and Chen (2011) found no difference between the *in vitro* dynamic mechanical response of brain matter in different animals (porcine, bovine and caprine), different breeds and different genders. Because of the similarities of porcine and human brain tissue, it is convenient to use porcine brain tissue for material characterization and to use these material parameters in human finite element head models.

On a microscopic scale, the brain is made up of billions of cells that interconnect and communicate (Nicolle et al., 2004). One of the most pervasive types of injury following even a minor trauma is damage to the nerve cell's axon through shearing during DAI. DAI in animals and human has been hypothesized to occur at macroscopic shear strains of 10% – 50% and strain rates of approximately 10 – 50/s (Margulies et al., 1990; Meaney and Thibault, 1990). Several studies have been conducted to determine the range of strain and strain rates associated with DAI. Bain and Meaney (2000) investigated *in vivo*, tissue-level, mechanical thresholds for axonal injury by developing a correlation between the strains experienced in the guinea pig optic nerve and morphological and functional injury. The threshold strains predicted for injury ranged from 0.13 – 0.34. Similarly, Pfister et al., (2003) developed a uniaxial stretching device to study axonal injury and neural cell death by applying strains within the range of 20%–70% and strain rates within the range of 20 – 90/s to create mild to severe axonal injuries. Bayly et al., (2006) carried out *in vivo* rapid indentation of rat brain to determine strain fields using harmonic phase analysis and tagged MR images. Values of maximum principal strains > 0.20 and strain rates > 40/s were observed in several animals exposed to 2mm impacts of 21 ms duration. Studies conducted by Morrison et al., (2000; 2003; 2006) also suggested that the brain cells are significantly damaged at strains > 0.10 and strain rates > 10/s.

In this study, mechanical properties of porcine brain tissue have been determined by performing tests at 30, 60 and 90/s strain rates up to 30% strain. The loading rates in the present study approximately cover the range of strain rates as revealed during TBI investigations by various research groups (Bain and Meaney, 2000; Bayly et al., 2006;

Margulies et al., 1990; Meaney and Thibault, 1990; Morrison et al., 2000; 2003; 2006; Pfister et al., 2003). The challenge with these tests was to attain uniform velocity during the tension phase of the experiments. Therefore a High Rate Tension Device (HRTD) was designed to achieve uniform velocity at dynamic loading velocities during extension of brain tissue. To fully characterize the behavior of brain tissue, material parameters have been determined by fitting isotropic one-term Ogden, Fung and Gent models. Force relaxation tests in tension were also conducted at various strain magnitudes (10% - 60% strain) with an average rise time of 24 ms. Relaxation data was used to estimate time dependent parameters. Numerical simulations were performed in ABAQUS Explicit/6.9 using material parameters from the one-term Ogden model. This study may provide new insight into the behavior of brain tissue under dynamic impact conditions, which would assist in developing effective brain injury criteria and adopting efficient countermeasures against TBI.

## 2 Materials and Method

### 2.1 Specimen Preparation

Ten fresh porcine brains from approximately six month old pigs were collected approximately 6 h after death from a local slaughter house and tested within 4 h, which was consistent with previous work (Estes and McElhaney, 1970; Miller and Chinzei, 1997; Tamura et al., 2007). Each brain was preserved in a physiological saline solution (0.9% NaCl /154 mmol/L) at 4 to 5$^{o}$C during transportation. All samples were prepared and tested at room temperature ~ 22 $^{o}$C. The dura and arachnoid were removed and the cerebral hemispheres were first split into right and left halves by cutting through the corpus callosum. As shown in Fig. 1, one half of the cerebral hemisphere was cut in the coronal plane to extract two coronal slices. The samples were then inserted in a cylindrical metal disk with 15.2 mm internal diameter and 10.1 mm thickness; any excessive brain portion was then removed with a surgical scalpel. The scalpel was dipped in a saline solution before cutting brain tissue to avoid sticking. The actual diameter and height of specimens measured before testing were 15.1±0.1 mm and 10.0±0.1 mm (mean ± SD). Miller and Chinzei (2002) also used a sample height of 10.0 mm during tension tests at quasistatic velocities (0.005, 5.0 and 500 mm/min). The time elapsed between harvesting of the first and the last specimens from each brain was 16 ~ 20 minutes. Physiological saline solution was applied frequently to specimens during cutting and before testing in order to prevent dehydration. The specimens were not all excised simultaneously, rather each specimen was tested first and then another specimen was extracted from the cerebral hemisphere. This procedure was important to prevent the tissue from losing some of its stiffness and to prevent dehydration, and thus contributed towards repeatability in the experimentation.

## 2.2 Experimental Setup

A *High Rate Tension Device (HRTD)* was designed and calibrated to perform tensile tests at strain rates of 30, 60 and 90/s to characterize the behavior of brain tissue under TBI conditions (Rashid et al., 2012a). However, for the present work the experimental setup was modified to perform tensile tests with thicker samples (≤ 14.0 mm), as shown in Fig. 2. The major components of the apparatus include a *servo motor controlled programmable electronic actuator* with a stroke length of 700 mm and a maximum velocity of 1500 mm/s (*LEFB32T-700,* SMC Pneumatics), two *5 N load cells* (GSO series -5 to +5 N, Transducer Techniques) with a rated output of 1.46 mV/V nominal and a safe overload of 150% of rated output, a *Linear Variable Displacement Transducer* (ACT1000 LVDT, RDP Electronics) with a range ± 25 mm, linearity ± 0.25 percent of full range output were used. An integrated single-supply instrumentation amplifier (AD 623 G = 100, Analog Devices) with built-in single pole low-pass filter having a cut-off frequency of 10 kHz was used. The output of the amplifier was passed through a second single pole low-pass filter with a cut-off frequency of 16 kHz. The amplified signal was analyzed through a data acquisition system with a sampling frequency of 10 kHz.

The force (N) and displacement (mm) data were recorded against time (s) for the tissue experiencing 30% strain. High speed image recording of brain tissue during tension tests was done at a frame rate of 3906 fps with 640 x 480 resolutions by using a high speed digital camera (Phantom V5.1, CMOS 10 bit Sensor). The images were examined to inspect that the cylindrical brain samples were uniformly deformed and the faces of the specimen were firmly bonded to the moving and stationary platens during extension of the brain specimen, as shown in Fig. 3.

## 2.3 Calibration to Achieve Uniform Velocity

Calibration of the HRTD was essential in order to ensure uniform velocity during extension of brain tissue at each strain rate. Two main contributing factors for the non-uniform velocity were the deceleration of the electronic actuator when it is approaching the end of the stroke and the opposing forces acting against the striking mechanism. Therefore, the striking mechanism (see Fig. 2) was designed and adjusted to ensure that it impacts on the tension pin approximately 150 mm before the actuator comes to a complete stop. The *striker* impact generates backward thrust, which is fully absorbed by the spring mounted on the actuator guide rod to prevent any damage to the *programmable servo motor*. Moreover, the actual actuator velocity was kept higher than the required (theoretically calculated) velocity to overcome the opposing forces acting against the striking mechanism (LVDT probe and sliding components). During the calibration process, the actuator was run several times to achieve uniform velocity. Experimentation with the brain tissue specimen started after the setup was fully calibrated at a particular velocity.

### 2.4  Specimen Mounting and Bonding Procedure

The surfaces of the platens were first covered with a masking tape substrate to which a thin layer of surgical glue (Cyanoacrylate, Low-viscosity Z105880–1EA, Sigma-Aldrich) was applied. The prepared cylindrical specimen of tissue was then placed on the lower platen. The top platen, which was attached to the 5 N load cell, was then lowered slowly so as to just touch the top surface of the specimen. One minute settling time was sufficient to ensure proper adhesion of the specimen to the top and lower platens. This procedure provided excellent attachment of the tissue to the platens. A calibrating metal disk of 10.0 mm thickness was also used to confirm the required distance between the platens before the start of experimentation. During tensile tests, excellent bonding was achieved (no slip boundary condition) at the brain/ platen interface due to the application of a thin layer of surgical glue (Cyanoacrylate, Low-viscosity Z105880–1EA, Sigma-Aldrich), as shown in Fig. 4. However, due to no slip boundary conditions, inhomogeneous deformation of the brain tissue was also observed at the edges of the brain/platen interface.

A high speed camera was used to monitor and record all tension tests, as discussed in Section 2.2. The images were used to confirm proper adhesion of brain tissue to the platens during the extension phase.

### 2.5  Tension Tests

Tension tests were performed up to 30% strain. The velocity of the platen producing extension in the brain tissue was adjusted to 300, 600 and 900 mm/s to attain approximate strain rates of 30, 60 and 90/s, respectively. Four cylindrical specimens containing mixed white and gray matter were extracted from the coronal plane (see Fig. 1) from each brain (40 specimens from 10 brains). The attainment of uniform velocity was also confirmed during the calibration process. No preconditioning was performed due to the extreme delicacy and tackiness of brain tissue. All tests were conducted at room temperature ~ 22 °C. A visible contraction of the cylindrical samples occurred immediately after they were removed from the brains, revealing the presence of residual stresses *in-vivo*. When measuring the dimensions of the specimens, it was noted that the nominal dimensions were reached after a few minutes; it was at this stage that testing commenced.

## 2.6  Force Relaxation in Tension

A separate set of force relaxation tests in tension was performed on cylindrical specimens (10.0 ± 0.1 mm thick and 15.0 ± 0.1 mm diameter). Here, 64 specimens were extracted from 8 brains (4 samples from each cerebral hemisphere). Ten force relaxation tests were performed at each of 10%, 20%, 30%, 40%, 50% and 60% strain in order to investigate the response of brain tissue to a step-like strain at variable strain magnitudes. The specimens for the relaxation tests were mounted according to the procedure described in Section 2.4. The specimens were stretched at various loading levels (300 – 700 mm/s) and held at the same position while measuring the relaxation force. The average rise time measured from the force relaxation experiments was approximately 24 milliseconds (ms). Force vs. time data was recorded for up to 1.6 s. Force relaxation experiments at dynamic strain rates are very important for the determination of time dependent parameters such as $\tau_k$, the characteristic relaxation times, and $g_k$, the relaxation coefficients. These are required to simulate impact conditions related to TBI. These parameters can then be used directly in a suitable constitutive model for the determination of stress, taking into account the strain rate dependency of the material.

# 3  Constitutive Models

## 3.1  Preliminaries

In general, an isotropic hyperelastic incompressible material is characterized by a strain-energy density function *W* which is a function of two principal strain invariants only: *W = W(I₁, I₂)*, where *I₁* and *I₂* and are defined as (Ogden, 1997),

$$I_1 = \lambda_1^2 + \lambda_2^2 + \lambda_3^2 \tag{1}$$

$$I_2 = \lambda_1^2\lambda_2^2 + \lambda_1^2\lambda_3^2 + \lambda_2^2\lambda_3^2 \tag{2}$$

Here, $\lambda_1^2, \lambda_2^2, \lambda_3^2$ are the squares of the principal stretch ratios, linked by the relationship $\lambda_1 \lambda_2 \lambda_3 = 1$, due to incompressibility.

It was not possible to achieve homogeneous deformation conditions due to the bonding of brain tissue (no slip conditions) at the platen/brain interfaces during tension tests; this is regarded as a practical limitation of our experimental protocol. Nevertheless, an effort was made to achieve approximately uniform contraction of the brain tissue with a specimen thickness of 10.0 ± 0.1 mm (mean ± SD) up to 30% strain. Therefore, we have assumed homogenous deformation of the brain tissue under tension. Then the Eulerian and Lagrangian principal axes of strain and stress are aligned with the direction of tension, $x_1$, say, and with any two orthogonal axes (lateral) $x_2$, $x_3$, say. Due to symmetry and incompressibility, the stretch ratios are now of the form

$$\lambda_1 = \lambda \text{ and } \lambda_2 = \lambda_3 = \frac{1}{\sqrt{\lambda}} \tag{3}$$

where $\lambda \geq 1$ is the stretch ratio in the direction of tension. Also, Eqs. (1) and (2) give

$$I_1 = \lambda^2 + 2\lambda^{-1} \text{ and } I_2 = \lambda^{-2} + 2\lambda, \tag{4}$$

so that $W$ is now a function of $\lambda$ only. During the experimental tension tests, the principal stretch ratio $\lambda$ was calculated from the measure of the elongation $e$ using equation: $\lambda = 1 + e$. The nominal/Lagrange stress component along the direction of tension $S_{11}$ was evaluated as $S_{11} \equiv F/A$, where $F$ is the tension force, as measured in Newtons by the load cell, and $A$ is the area of a cross section of the sample in its undeformed state. The experimentally measured nominal stress was then compared to the predictions of the hyperelastic models from the relation (Ogden, 1997),

$$S_{11} = \frac{d\tilde{W}}{d\lambda}, \text{ where } \tilde{W}(\lambda) \equiv W(\lambda^2 + 2\lambda^{-1}, \lambda^{-2} + 2\lambda), \tag{5}$$

and the material parameters were adjusted to give good curve fitting.

### 3.2 Fung Strain Energy Function

The Fung isotropic strain energy (Fung, 1967; Fung et al., 1979) is often used for the modelling of soft biological tissues in tension. It depends on the first strain invariant only:

$$W = \frac{\mu}{2b} \left[ e^{b(I_1 - 3)} - 1 \right] \tag{6}$$

It yields the following nominal stress component $S_{11}$ along the $x_1$-axis,

$$S_{11} = \mu e^{b(\lambda^2 + 2\lambda^{-1} - 3)}(\lambda - \lambda^{-2}) \qquad (7)$$

Here $\mu > 0$ (infinitesimal shear modulus) and $b > 0$ (stiffening parameter) are the two constant material parameters to be adjusted in the curve-fitting exercise.

### 3.3 Gent Strain Energy Function

The Gent isotropic strain energy (Gent, 1996) describes rapidly strain-stiffening materials in a very satisfying way. It also depends on the first strain invariant only, as

$$W(I_1) = -\frac{\mu}{2} J_m \ln\left(1 - \frac{I_1 - 3}{J_m}\right) \qquad (8)$$

It yields the following nominal stress $S_{11}$ along the $x_1$-axis

$$S_{11} = \frac{\mu J_m}{J_m - \lambda^2 - 2\lambda^{-1} + 3}(\lambda - \lambda^{-2}) \qquad (9)$$

Here $\mu > 0$ (infinitesimal shear modulus) and $J_m > 0$ are two constant material parameters to be optimized in the fitting exercise.

### 3.4 Ogden Strain Energy Function

The Ogden model (1972) has been used in the past to describe the nonlinear mechanical behavior of the brain, as well as of other nonlinear soft tissues (Brittany and Margulies, 2006; Lin et al., 2008; Miller and Chinzei, 2002; Prange and Margulies, 2002; Velardi et al., 2006). Soft biological tissue is often modeled well by the Ogden formulation and most of the mechanical test data available for brain tissue in the literature are fitted with an Ogden hyperelastic function. The one-term Ogden hyperelastic function is given by

$$W = \frac{2\mu}{\alpha^2}\left(\lambda_1^\alpha + \lambda_2^\alpha + \lambda_3^\alpha - 3\right) \qquad (10)$$

It yields the following nominal stress $S_{11}$

$$S_{11} = \frac{2\mu}{\alpha}\left\{\lambda^{\alpha-1} - \lambda^{-\left(\frac{\alpha}{2}+1\right)}\right\} \qquad (11)$$

Here $\mu > 0$ is the infinitesimal shear modulus, and $\alpha$ is a stiffening parameter.

# 4 Results

## 4.1 Tensile Experiments

Ten tensile tests on cylindrical specimens were performed at each strain rate of 30, 60 and 90/s up to 30% strain in order to analyze experimental repeatability and behavior of tissue at a particular loading velocity, as shown in Fig. 5. Force (N) and displacement (mm) data measured directly through the four channel data acquisition system (Handyscope, HS4) at a sampling frequency of 10 kHz was converted to engineering stress (kPa) – time (s) for each strain rate. It was observed that the tissue stiffness increases with the increase in loading velocity, indicating the strong stress – strain rate dependence of brain tissue. The maximum engineering stress at 30% strain at strain rates of 30, 60 and 90/s are 3.1±0.49 kPa, 4.3±0.86 kPa, 6.5±0.76 kPa (mean ± SD), respectively.

## 4.2 Fitting of Constitutive Models

Experimental stress values and corresponding stretch ratios of each anatomical region were used to perform a non-linear least-square fit of the parameters for three common hyperelastic constitutive models. The constitutive models were fitted to the stress-stretch data for strains up to 30%.

The fitting was performed using the *lsqcurvefit.m* function in MATLAB, and the quality of fit for each model was assessed based on the goodness of the Coefficient of determination $R^2 = \frac{S_t - S_r}{S_t}$, where $S_t$ = the total sum of the squares of the residuals between the data points and the mean and $S_r$ = sum of the squares of the residuals around the regression line. The fitting of hyperelastic models has been comprehensively covered by Ogden et al., (2004). The average engineering stress – stretch curves at each loading rate, as shown in Fig. 5, were used for fitting to hyperelastic isotropic constitutive models (Fung, Gent and Ogden models) given by Eqs. (7), (9) and (11). Excellent fit is achieved for all models (coefficient of determination: $0.9980 < R^2 \leq 0.9999$) and the resulting theoretical curves are indistinguishable from each other as shown in Fig. 6.

The material parameters were derived after fitting strain energy functions to each experimental engineering stress – stretch profile shown in Fig. 5. The mean and SD were calculated for all the material parameters. All best fit material parameters ($\mu$, $\alpha$, $b$, $J_m$) derived at each strain rate are summarized in Table 1.

### 4.3 Elastic Moduli of Brain Tissue

The elastic moduli at each strain rate were also calculated from the experimental data shown in Fig. 5. The elastic moduli $E_1$, $E_2$, and $E_3$ calculated from the tangent to the stress – strain curve corresponded to the strain ranges of 0 – 0.1, 0.1 – 0.2 and 0.2 – 0.3, respectively. The estimated elastic moduli at each strain rate are summarized in Table 2. There is a consistent rise in moduli with increasing strain ranges and with increasing strain rates.

### 4.4 Force Relaxation Experiments

The cylindrical specimens were stretched at various strain levels (10% - 60% strain) and held at the same position to record the relaxation force. The step-like loading of brain tissue in tension during relaxation at various strain levels is shown Fig. 7. Since the brain tissue relaxed on the time scale of the ramp, it was also necessary to include the ramp loading phase in each relaxation test. The standard deviation (SD) mentioned against each strain level is at the peak force (N), as shown in Fig. 7 (a) and (b). The peak force relaxed by approximately 83% (average of 10% to 60% strain) up to 0.1 s (Fig. 7 (c)), and then it continuously decreased gradually up to 0.2 s. The dramatic decrease in force reveals the highly viscoelastic nature of brain tissue.

### 4.5 Ogden-based Hyper-viscoelastic Model

Many nonlinear viscoelastic models have been formulated, but Fung's theory (Fung, 1993) of quasi-linear viscoelasticity (QLV) is probably the most widely used due to its relative simplicity. To account for the time-dependent mechanical properties of brain tissue, the stress-strain relationship is expressed as a single hereditary integral (others have used a similar approach, e.g., Elkin et al., 2011; Finan et al., 2012; Miller and Chinzei, 2002),

$$S_{11} = \frac{2\mu}{\alpha} \int_0^t G(t-\tau) \frac{d}{d\tau}\left( \lambda^{\alpha-1} - \lambda^{-\left(\frac{\alpha}{2}+1\right)} \right) d\tau \tag{12}$$

Here, $S_{11}$ is the nominal tensile stress component, $\mu$ is the initial shear modulus in the undeformed state, and $\alpha$ is a stiffening parameter derived from the one-term Ogden model (Eq 10 and 11), where $\mu > 0$. The relaxation function $G(t)$ is defined in terms of Prony series parameters:

$$G(t) = \left[1 - \sum_{k=1}^{n} g_k (1 - e^{-t/\tau_k})\right] \quad (13)$$

$\tau_k$ are the characteristic relaxation times, and $g_k$ are the relaxation coefficients. In order to estimate material parameters with a physically meaningful interpretation, we solved Eq. 12 using Matlab functions.

### 4.6 Estimation of Viscoelastic Parameters

It was convenient to solve Eq. 12 in Matlab 6.9 by using the *gradient* and *conv* functions. The *gradient* function was used in order to determine the velocity vector $\frac{d}{d\tau}\left(\lambda^{\alpha-1} - \lambda^{-\frac{\alpha}{2}-1}\right)$ from the experimentally measured displacement, $\lambda$, and time, $\tau$. Also, a *conv* function was used to convolve the entire force history with velocity vector. The raw data (acquired at 10 kHz) was reduced before fitting. We estimated the Prony parameters ($g_1$ = 0.5663, $g_2$ = 0.3246, $\tau_1$ = 0.0350 s, $\tau_2$ = 0.0351 s) from a two-term relaxation function. The coefficients of the relaxation function were optimized using *nlinfit* and *lsqcurvefit* to minimize the error between the experimental stress data and Eq. (12). The initial shear modulus $\mu$, from the Ogden model should be independent of time and can be obtained from the nonlinear elastic response of the tissue at maximum strain rates.

## 5 Finite Element Analysis

### 5.1 Numerical and Experimental Results

The inhomogeneous deformation of the brain tissue up to 30% strain, particularly near the platen ends, due to no slip boundary conditions was observed during extension of the brain tissue at various loading conditions. Theoretical model developments as discussed in Section 3, do not encompass any inhomogeneous deformation of the brain tissue, thus theoretical deviation from the actual experimentation is expected. Therefore, at this stage, it was necessary to validate the predictable behavior of the theoretical model (using material parameters of one-term Ogden strain energy function, see Table 1) of the brain tissue against a numerical model and measured experimental data. Numerical simulations were performed by applying various boundary conditions using ABAQUS 6.9/Explicit to mimic experimental conditions. The mass density 1040 kg/m$^3$ and hexagonal C3D8R elements were used for the brain part. Velardi et al. (2006) used a mass density of 1039 and 1036 kg/m$^3$ for the white matter and the gray matter, respectively. Moreover, mass density was also estimated experimentally during this study and found to be 1040±1 kg/m$^3$. The bottom surface of the cylindrical specimen (15.0 mm diameter and 10.0 mm thick) was stretched in order to achieve 30% strain, whereas the top surface was constrained in all directions. Before numerical simulations, a mesh convergence analysis was carried out by varying mesh density. The mesh was considered converged when there was a negligible change in the numerical solution (0.9%) with further mesh refinement. The total number of elements for the specimen was 3421 with average simulation time of 60 s.

Excellent agreement between the average experimentally measured engineering stresses and the numerically predicted engineering stresses (kPa) was achieved, as shown in Fig. 8 (a). Moreover, the experimentally measured force (N) was also validated against force determined numerically in a separate set of simulations. Excellent agreement between the average experimental force and the numerical force (N) was achieved, as shown in Fig. 8 (b). Figures 8 (c) and (d) show deformed states of the cylindrical specimen with stress and force contours, respectively, when stretched to achieve 30% strain.

The reaction forces are positive at the moving end and negative at the stationary end of the cylindrical specimen; however the magnitudes of these forces remain the same at both ends of the specimen, as shown in Fig. 8 (d). Similar simulations were also performed with material parameters estimated at a strain rate of 60 and 90/s. The patterns of stress and force contours were not significantly different from the results shown in Fig. 8 (c) and (d).

## 5.2   Statistical Analysis and Artificial Strain Energy

The experimental data shown in Fig. 8 (a) and (b) was analyzed statistically. Based on the one-way ANOVA test, there was no significant difference between the experimental and numerical engineering stresses: p = 0.6379, p = 0.9863 and p = 0.9254 at 30, 60 and 90/s strain rates, respectively, as shown in Fig. 9 (a). Similarly there was no significant difference between the experimental and numerical forces: p = 0.9190, p = 0.8793 and p = 0.8807 at 30, 60 and 90/s strain rates, respectively, as shown in Fig. 9 (b). The good agreement between the experimental and numerical results indicates that a one-term Ogden model is appropriate to characterize the behavior of the brain tissue up to 30% strain.

The *accumulated artificial strain energy* (ALLAE), used to control *hourglass deformation* during numerical simulations, was also analyzed. It was observed that ALLAE for the whole model as a percentage of the total strain energy was 0.74%, 0.85% and 0.75% for the numerical simulations performed at 30, 60 and 90/s strain rates, respectively. The low percentage of artificial strain energy (≤ 0.85%) observed during the simulations indicates that hourglassing is not a problem; this percentage could be reduced further using two dimensional numerical models.

# 6 Discussion

The characterization of brain tissue in tension at high strain rates is crucial in understanding the mechanism of TBI under impact conditions. In this research, the properties of porcine brain tissue in extension have been determined up to 30% strain at strain rates of 30, 60 and 90/s by using a custom-designed HRTD. The HRTD is specifically designed and calibrated for uniform velocity during the extension phase of brain tissue at strain rates ≤ 90/s. Force relaxation experiments in tension were also conducted at various strain magnitudes (10% - 60% strain) with an approximate rise time of 24 ms. The time dependent Prony parameters can be utilized for a hyperviscoelastic analysis of brain tissue under impact conditions. An excellent agreement between the experimental and numerical engineering stresses indicates that a one-term Ogden model is appropriate to model soft biological tissues in tension up to 30% strain.

The moduli of brain tissue determined by other research groups were also studied for comparison purposes. Morrison et al., (2003) assumed an Elastic modulus, $E$ of 10 kPa in their FE model to predict the strain field in a stretched culture of rat-brain tissue, in which the maximum strain and strain rates were 30% and 50/s respectively. This compares well with the mean Elastic modulus, $E_1$ (strain range: 0 – 0.1) estimated in this study, i.e., 11.68 ± 3.79 (kPa), as indicated in Table 2.

In this study, we have conducted tension tests at a room temperature of ~22 $^o$C. Miller and Chinzei (1997; 2002) also performed such tests at the same temperature. Hrapko et al., (2008) analyzed the difference between room temperature (approximately 23 $^o$C) and body temperature (approximately 37 $^o$C) conditions. The measured results were found to be clearly temperature dependent and the dynamic modulus $G^*$ at 23 $^o$C was approximately 35% higher than at 37 $^o$C. Stiffening of the samples occurred with decreasing temperature. Zhang et al., (2011) conducted tests on porcine brain tissue at high strain rates specifically to investigate stress – strain behavior at ice cold temperature and at 37 $^o$C. The estimated stresses at 37 $^o$C were 60 – 70% higher than at ice cold temperature, showing a stiffer response of brain tissue at higher temperature (37°C). Hrapko et al., (2008) showed a stiffer response of brain tissue at the lower temperature (23 °C). Based on these contradictory findings, there is a crucial need to further investigate the effects of temperature on brain tissue, although this is beyond the scope of the present work.

The primary difference between *in vivo* and *in vitro* mechanical behavior of brain tissue is mainly due to postmortem time and is considered to be the dominant cause for the large variation in results (Gefen and Margulies, 2004). However, the pressurized vasculature (during *in vivo* tests), loss of perfusion pressure (during *in vitro* tests) and inter-species variability have very little effect on the experimental results (Gefen and Margulies, 2004). Therefore, in the present study, all tests were completed within 4 h after collection from a slaughter house.

Another limitation of this study is the estimation of material parameters from the strain energy functions based on average mechanical properties (mixed white and gray matter) of the brain tissue; however, these results are still useful in modelling the approximate behaviour of brain tissue. Moreover, the average mechanical properties were also determined by Miller and Chinzei (1997; 2002). In previous studies, it was observed that the anatomical origin or location as well as the direction of excision of samples (superior – inferior and medial – lateral direction), had no significant effect on the results (Tamura et al., 2007) and similar observations were also reported by Donnelly and Medige (1997). Based on the research conducted by Prange and Margulies (2002), by using samples of size 55 mm x 10 mm and 1 mm thick, the gray matter showed no difference between the two orthogonal directions whereas the white matter showed significantly different behaviour. Similarly, in the case of directional properties across regions, comparisons were made between the corona radiata and corpus callosum. The corona radiata was significantly stiffer than the corpus callosum, and the white matter behavior was more anisotropic than gray matter. In our future research, we intend to characterize the mechanical behavior of white and gray matter as well as regional and directional properties at high strain rates (1, 10, 20, 30/s) using indentation methods.

## 7. Conclusions

The following results can be concluded from this study:
(i) – The estimated tensile engineering stress at 30% strain is 3.1±0.49 kPa, 4.3±0.86 kPa, 6.5±0.76 kPa (mean ± SD) at strain rates of 30, 60 and 90/s, respectively.

(ii) – One-term Fung, Gent and Ogden models provide excellent fitting to experimental data up to 30% strain (coefficient of determination: $0.9980 < R^2 \leq 0.9999$).

(iii) – Excellent agreement between the experimental and numerical engineering stresses indicates that the Ogden strain energy function is fully able to characterize the behavior of brain tissue in tension up to 30% strain. This model is readily available in ABAQUS/6.9 for numerical analysis.

(iv) – Time dependent Prony parameters can be utilized with the Ogden hyperelastic parameters to perform hyper-viscoelastic analysis of the brain tissue in ABAQUS/6.9.

(v) – The test apparatus (HRTD) can be used for soft biological tissues for the tension tests with confidence at strain rates $\leq 90$/s.

(vi) – The derived material parameters can be used in finite element simulations for forensic biomechanics investigations.

**Acknowledgements**    The authors thank John Gahan, Tony Dennis and Pat McNally of University College Dublin for their assistance in machining components and developing electronic circuits for the experimental setup. This work was supported for the first author by a Postgraduate Research Scholarship awarded by the Irish Research Council for Science, Engineering and Technology (IRCSET), Ireland.

Figure Captions

Fig. 1 – Locations of samples extracted (~15.1±0.1 mm diameter) from the porcine brain tissue, containing mixed white and gray matter.

Fig. 2 – Major components and schematic diagram of the complete test apparatus. Dashed and solid lines indicate inputs and outputs respectively from the electronic components. Components colored in red move down when striker impacts on the tension pin.

Fig. 3 – Extension of porcine brain specimen between the platens attached to the load cells. The extension results due to an impact of the striker on the tension pin at a specific velocity. The displacement is precisely controlled by the stopper plate.

Fig. 4 – Cylindrical brain specimen (10.0 ± 0.1 mm) fully stretched to achieve 30% strain. Arrows indicate downward movement of the lower platen producing stretch in the brain tissue at a specific strain rate.

Fig. 5 – Tensile tests performed on porcine brain tissue up to 30% strain at loading velocity of 300 mm/s (strain rate: 30/s), 600 mm/s (strain rate: 60/s) and 900 mm/s (strain rate: 90/s)

Fig. 6 – Excellent agreement between strain energy functions (Fung, Gent and Ogden models) and the experimental nominal stress is achieved at 30, 60 and 90/s strain rates ($p = 0.9990$)

Fig. 7 – (a) Relaxation force (N) at various strain magnitudes (b) Approximate rise time and relaxation force at various strain magnitudes (c) Average force and standard deviation, SD (dotted lines)

Fig. 8 – Simulations showing numerical stress and reaction force contours (a) agreement of experimental and numerical engineering stresses (kPa) at each strain rate (b) agreement of experimental and numerical forces (N) at each strain rate (c) cross-sectional view showing stress contours (Pa) at 30% strain (d) cross-sectional view showing reaction force contours (N) at 30% strain.

Fig. 9 – There is no significant difference between the numerical and experimental results (engineering stress (kPa) and Force (N)) based on one way ANOVA tests.

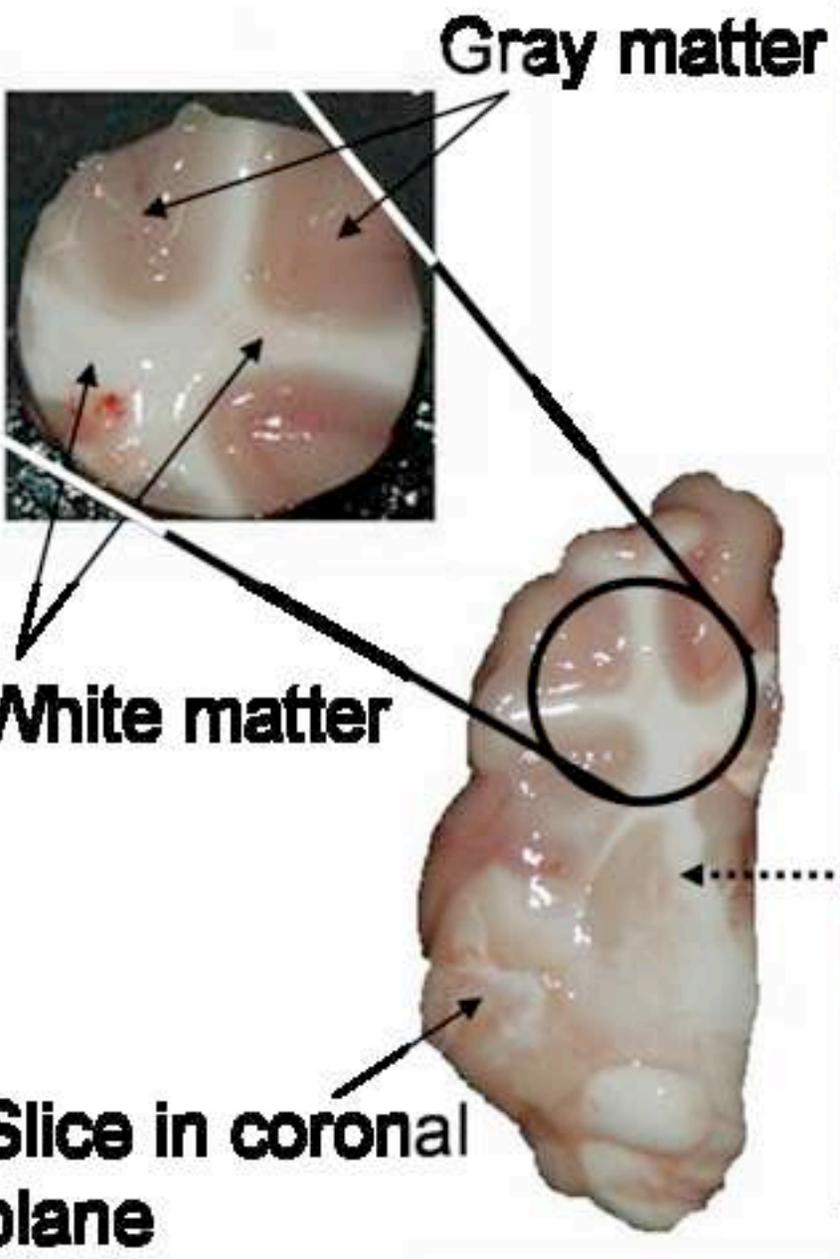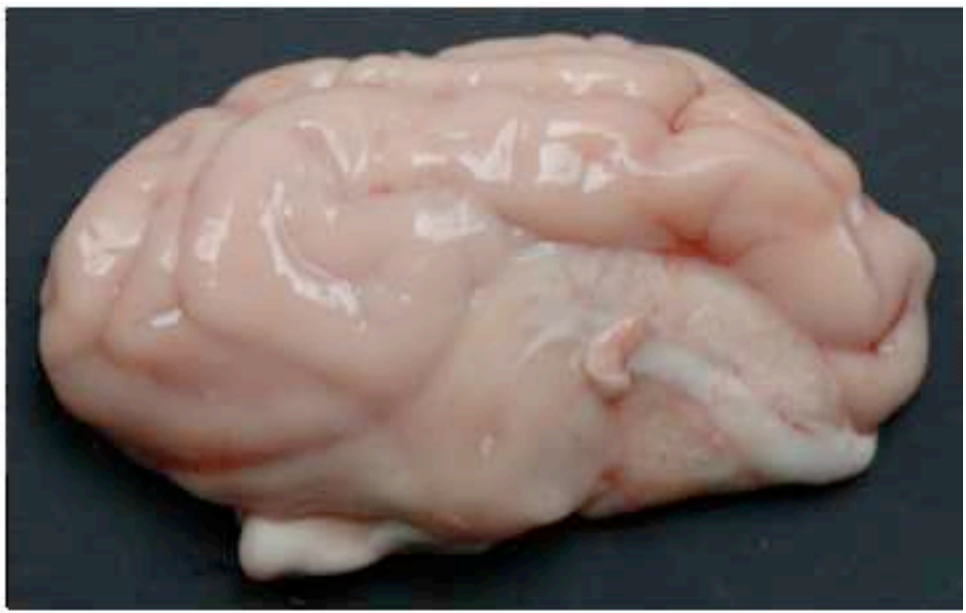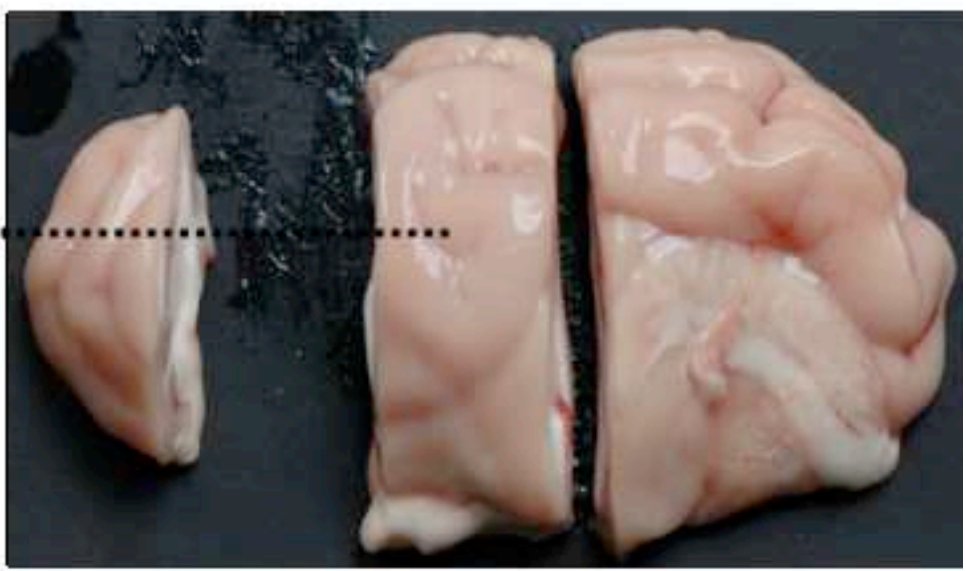

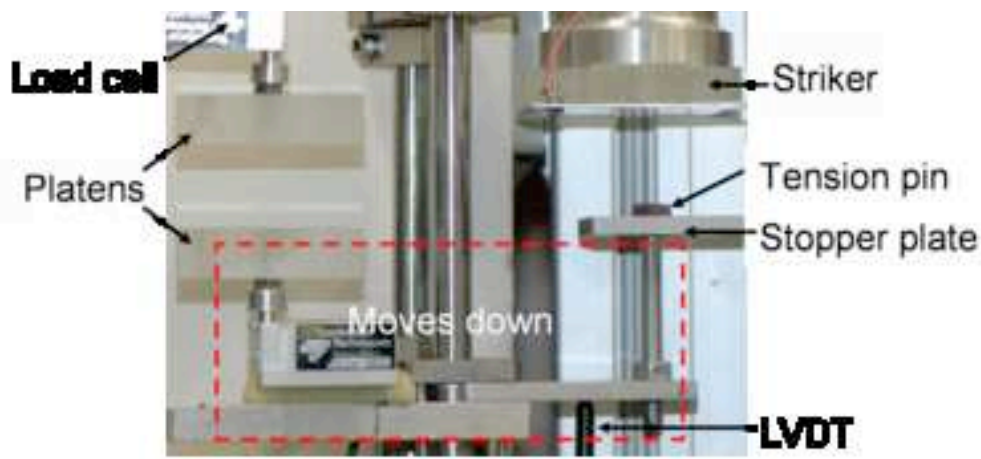
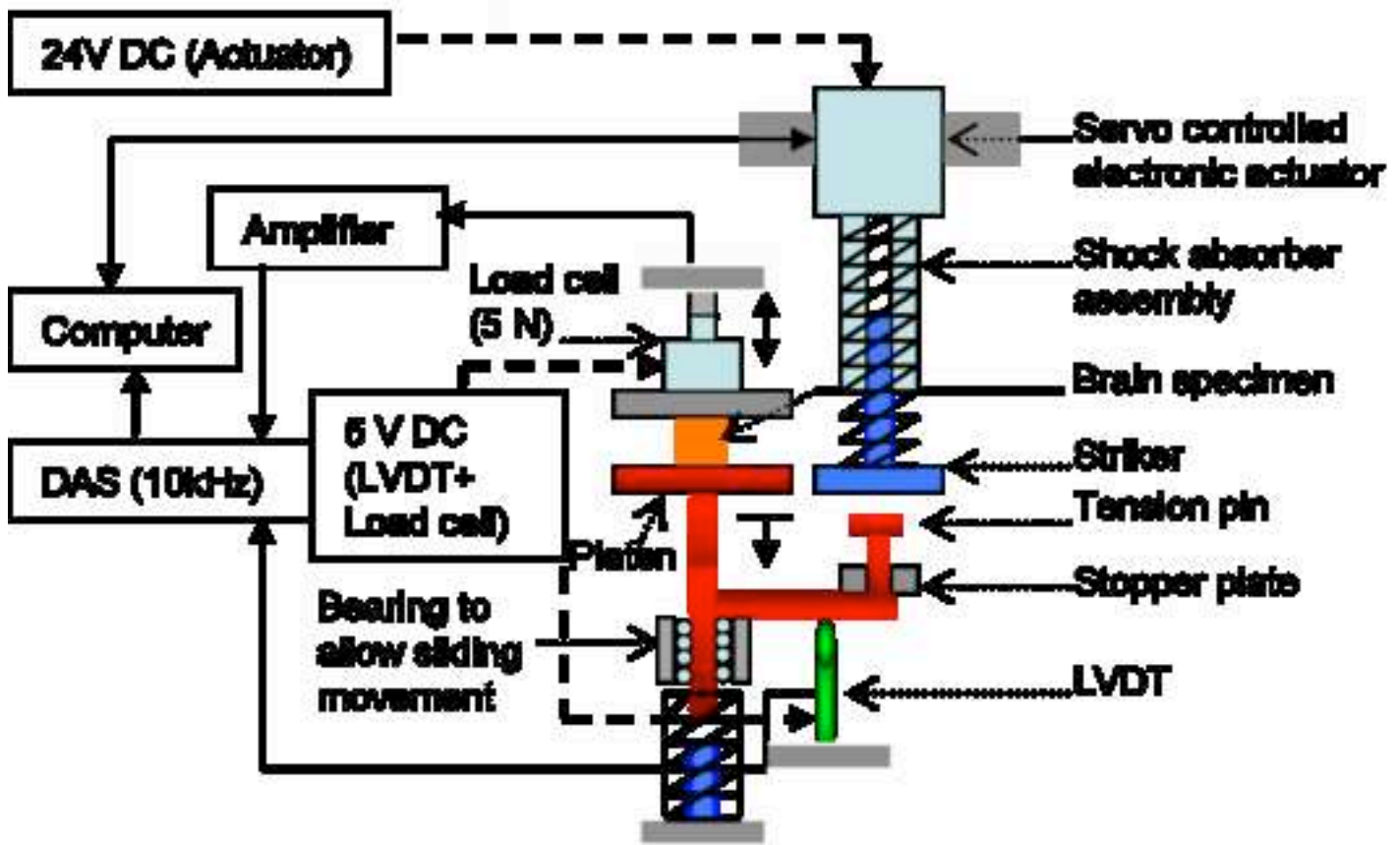

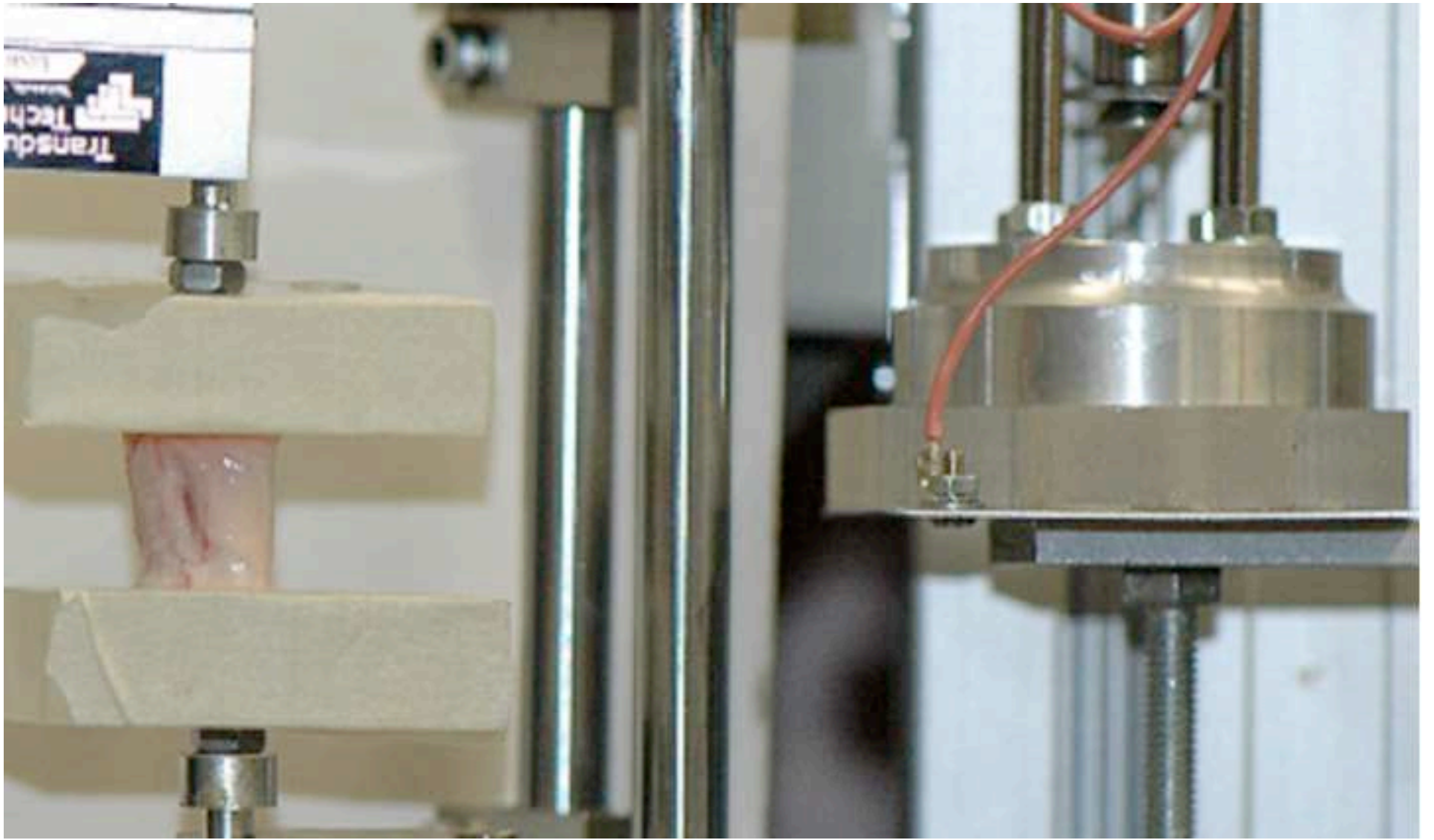

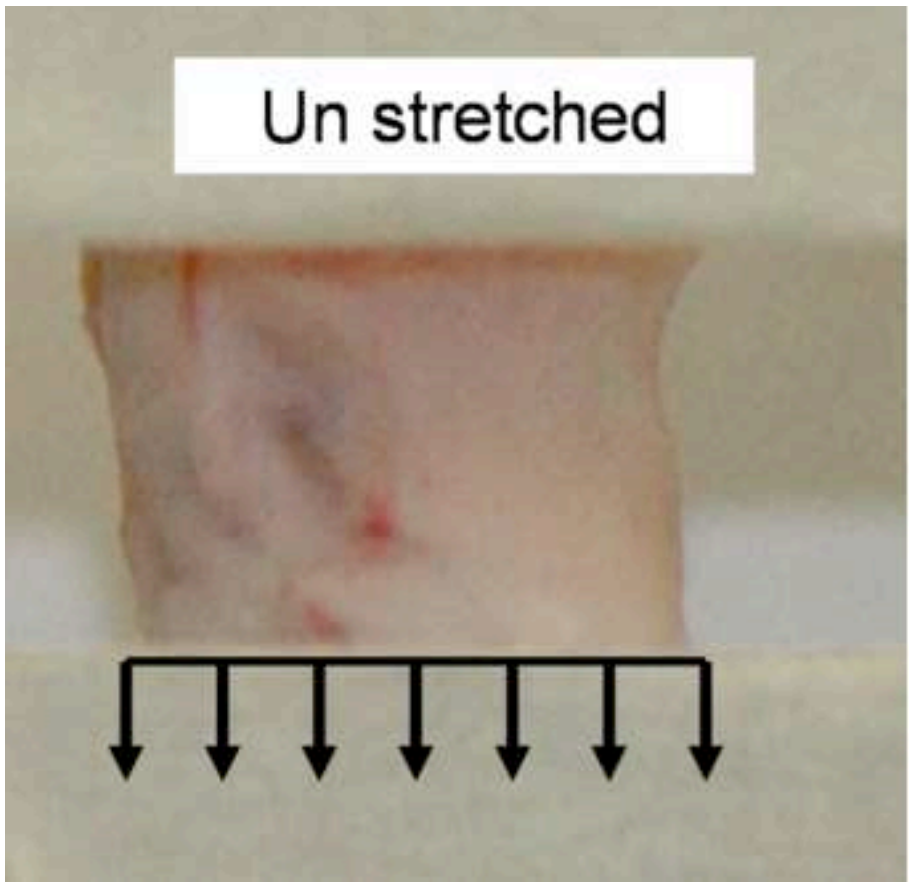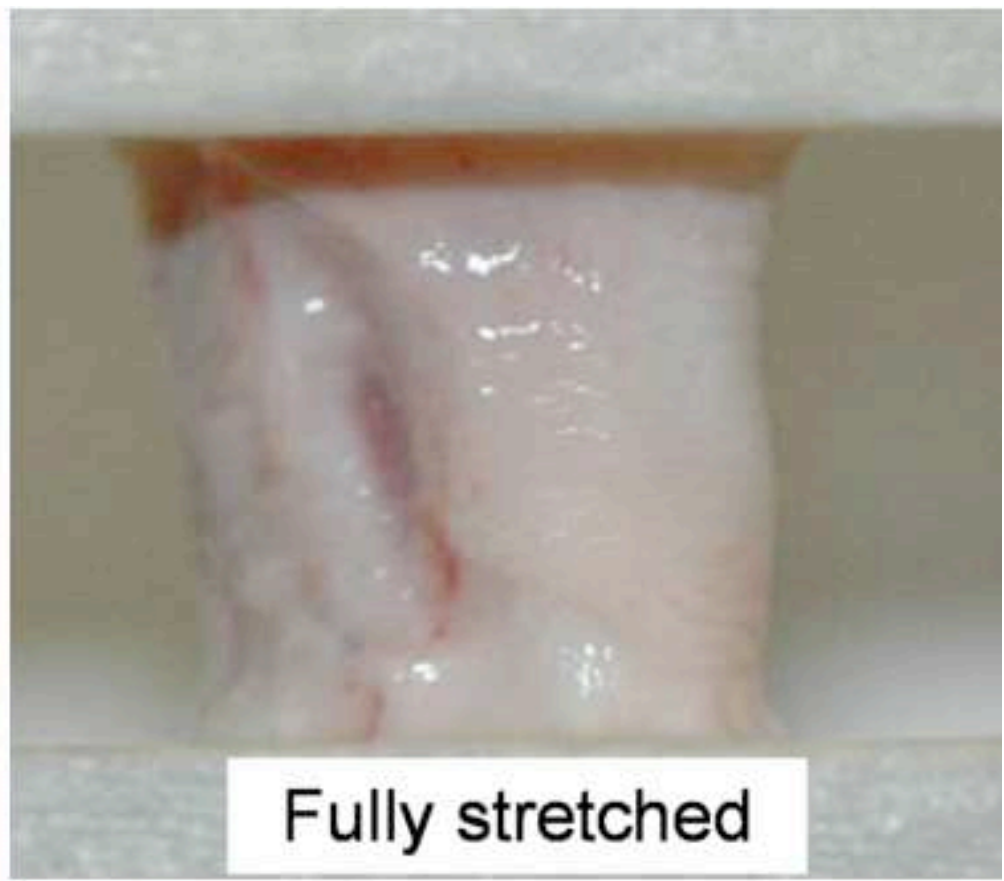

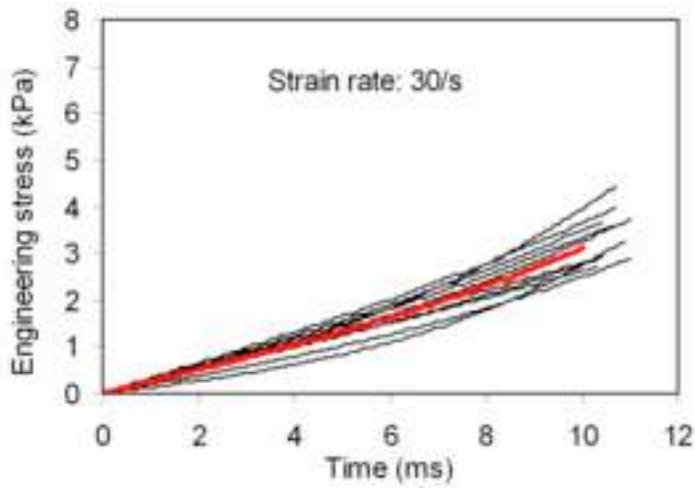
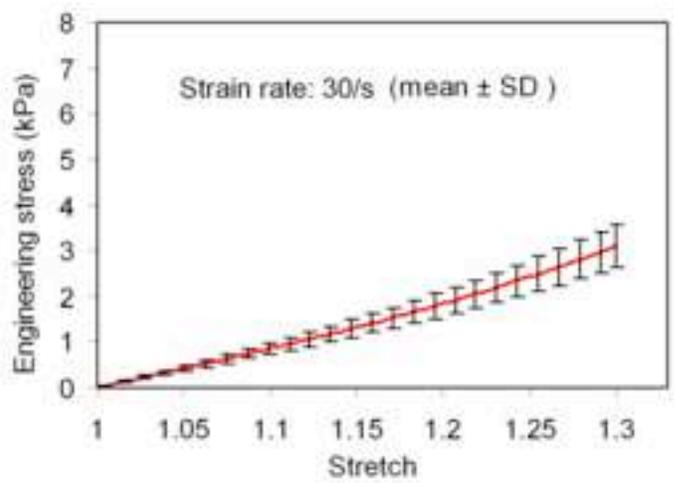
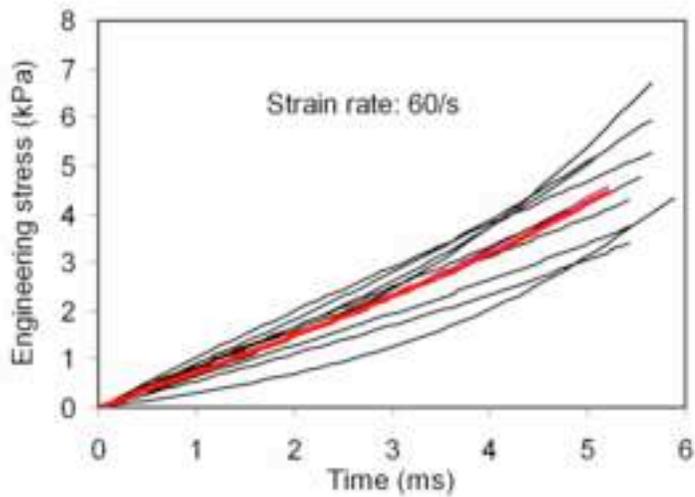
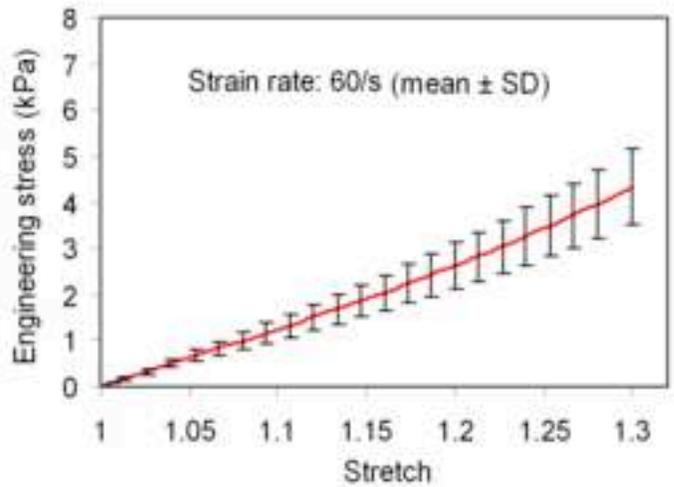
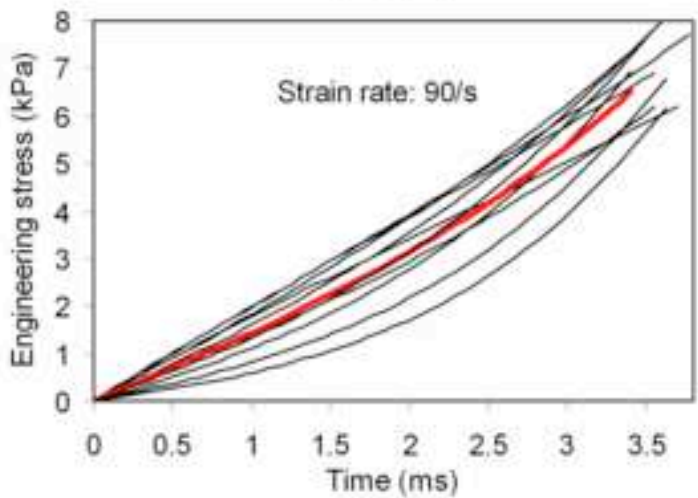
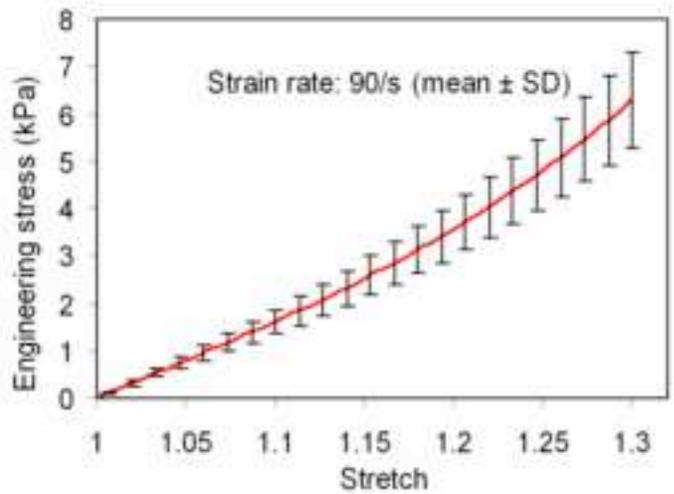

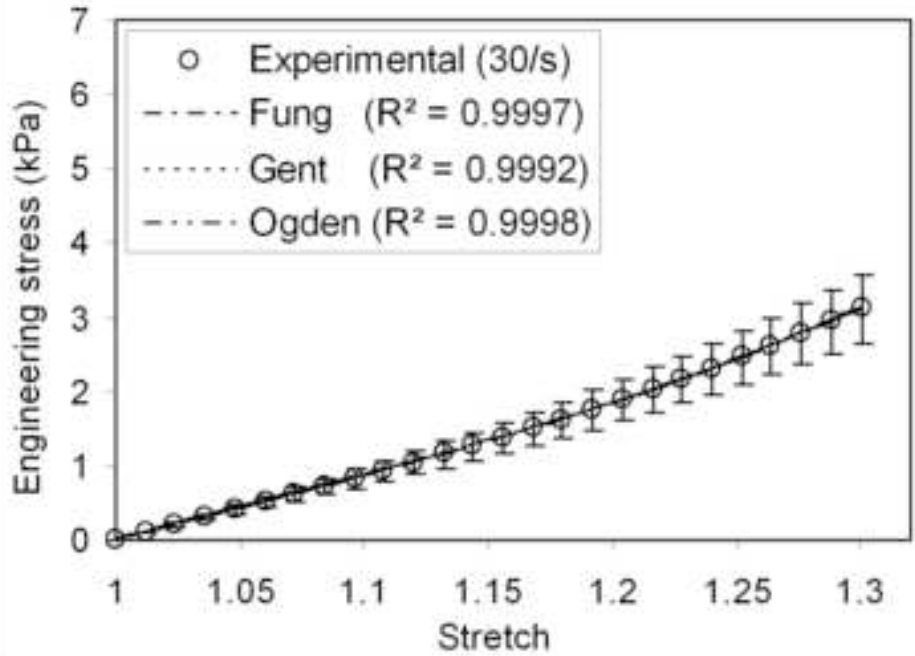
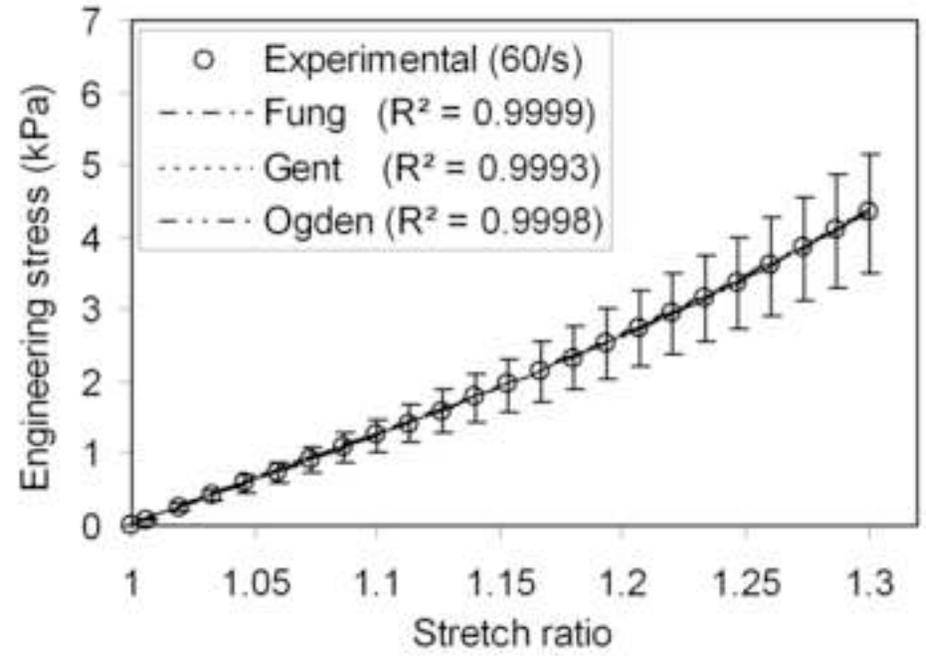
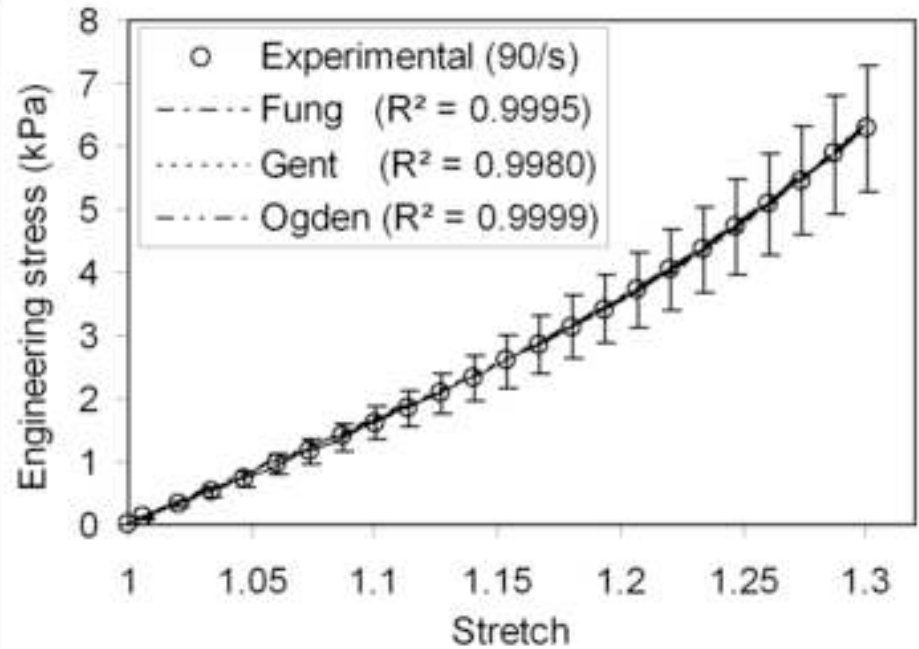

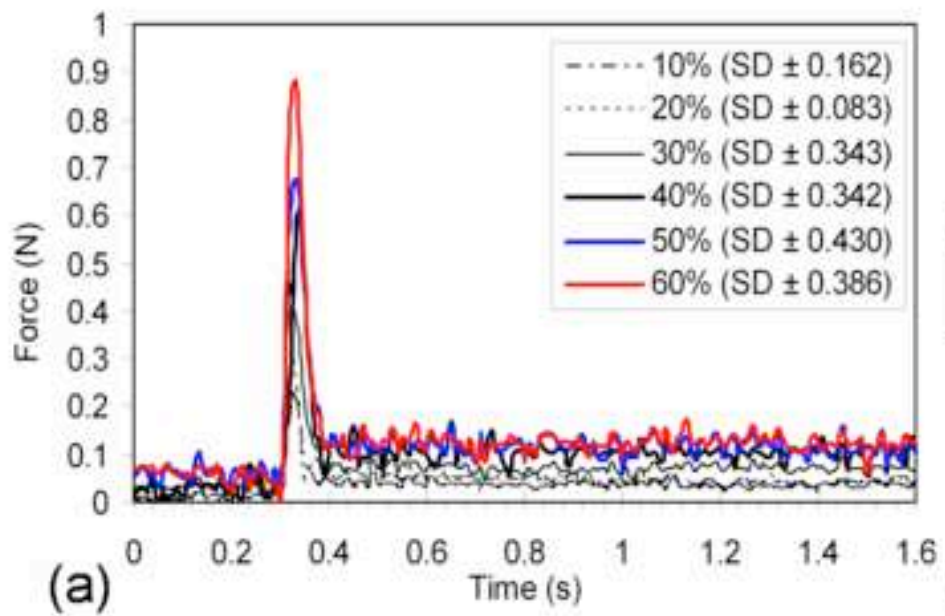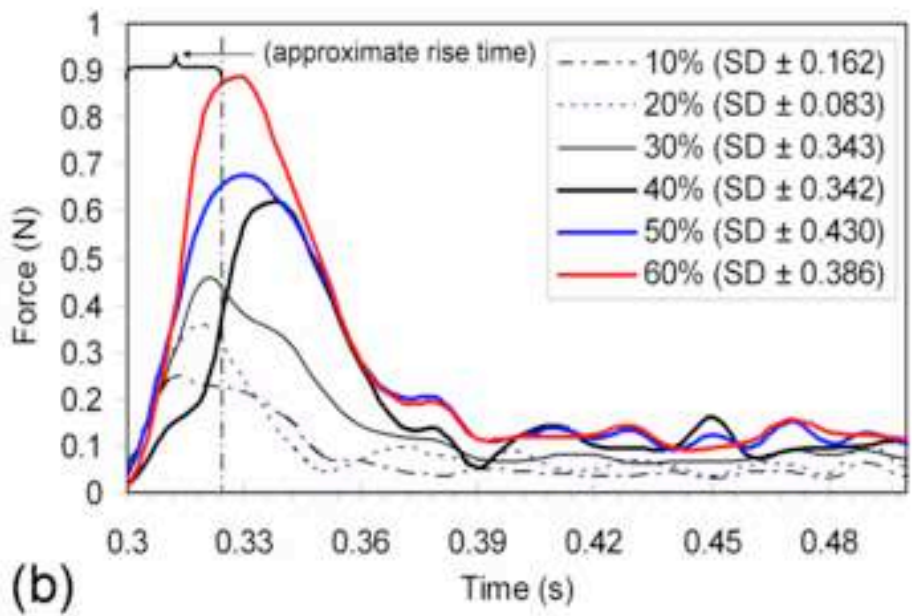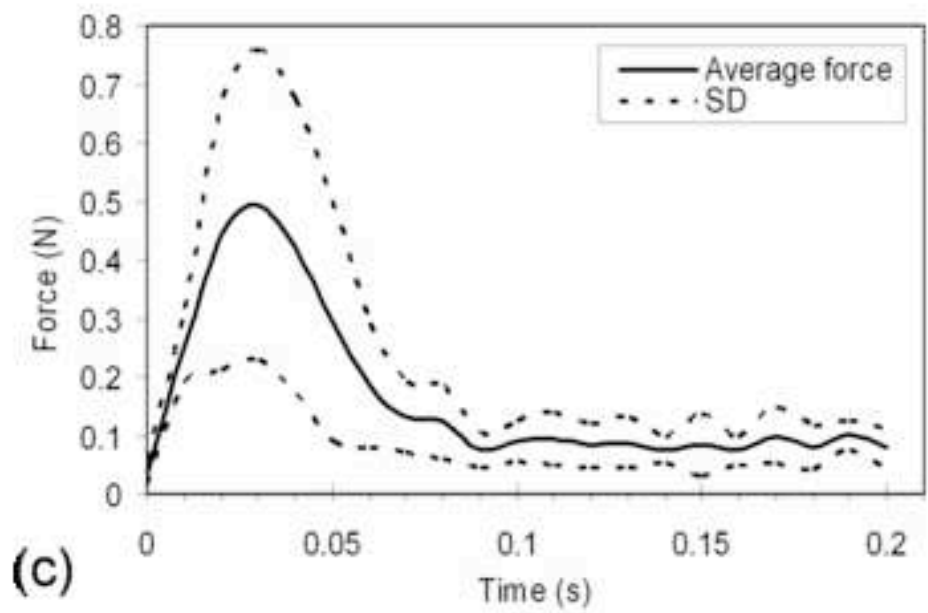

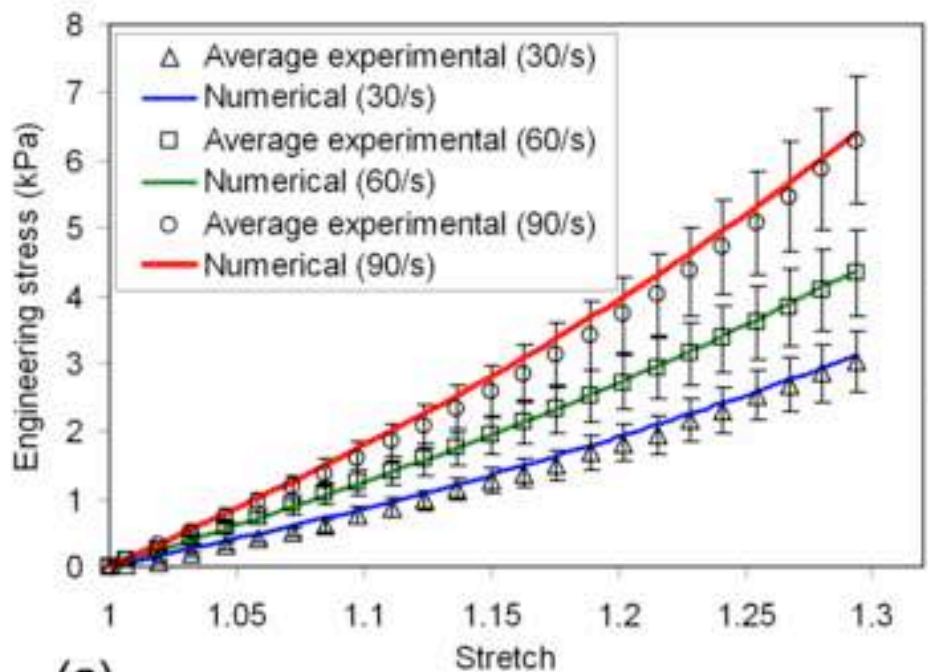
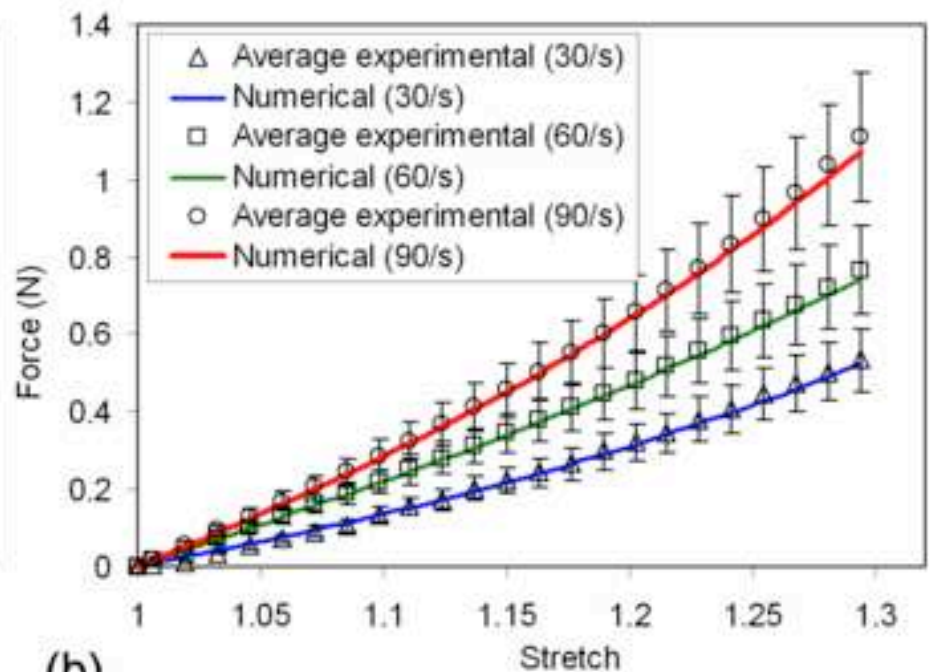
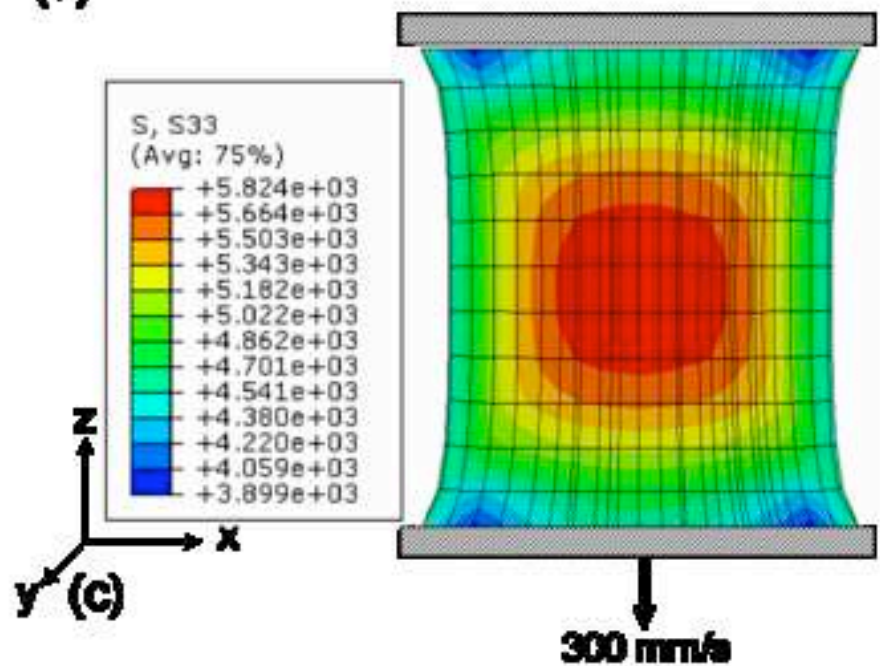
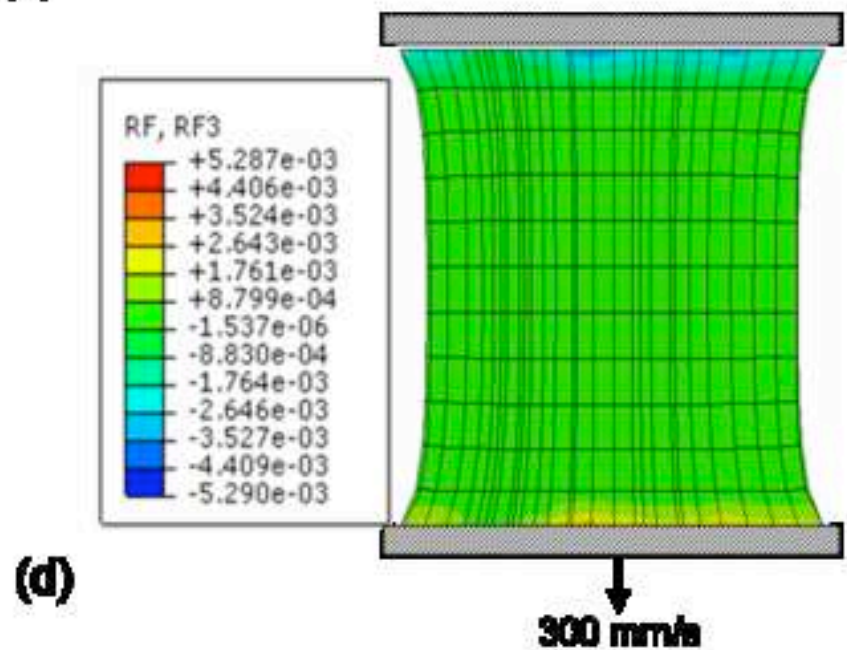

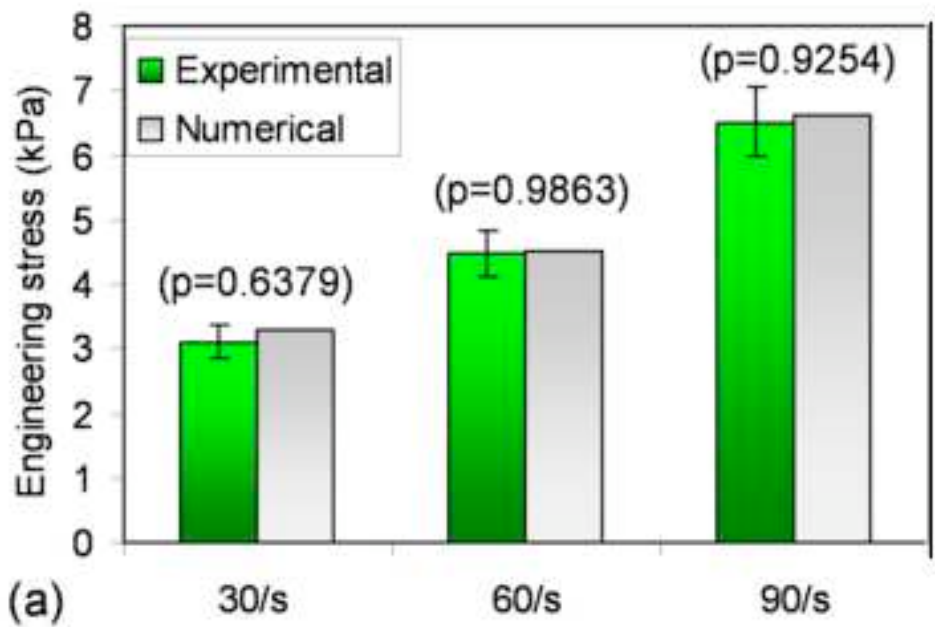 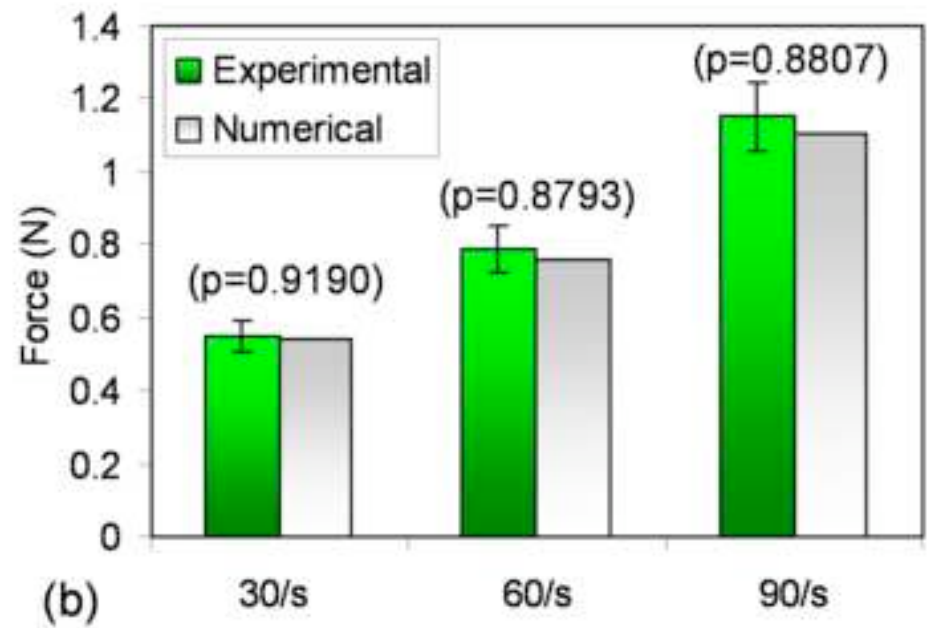

**Table**

| Table 1 – Material parameters derived after fitting of models to experimental data (all $\mu$ are in Pa (mean ± SD) and $\mu > 0$). | | | | | | |
|---|---|---|---|---|---|---|
| (1/s) | Fung model | | Gent model | | Ogden model | |
|  | $\mu$ | $b$ | $\mu$ | $J_m$ | $\mu$ | $\alpha$ |
| 30 | 3047 ± 643.0 | 1.68 ±0.88 | 3114 ± 611.2 | 0.86 ±0.32 | 2780 ± 657.4 | 6.0 ±1.72 |
|  | $R^2$ = 0.9997 ± 0.0002 | | $R^2$ = 0.9991 ± 0.0012 | | $R^2$ = 0.9997 ± 0.0001 | |
| 60 | 4458 ± 1174.6 | 1.5 ± 1.0 | 4548 ± 1127.4 | 1.15 ±0.71 | 4112± 1217.2 | 5.63±2.06 |
|  | $R^2$ = 0.9998 ± 0.0002 | | $R^2$ = 0.9993 ± 0.0009 | | $R^2$ = 0.9998 ± 0.0001 | |
| 90 | 5739 ± 1913.8 | 2.19 ± 1.47 | 5962 ± 1741.3 | 0.94±0.63 | 5160 ± 2045.2 | 6.95±2.85 |
|  | $R^2$ = 0.9995 ± 0.0005 | | $R^2$ = 0.9979 ± 0.0026 | | $R^2$ = 0.9999 ± 0.0001 | |

| Table 2 Elastic moduli of brain tissue at each strain rate (mean ± SD) | | | |
|---|---|---|---|
| Strain rate (1/s) | $E_1$ (kPa) | $E_2$ (kPa) | $E_3$ (kPa) |
| Strain range | 0 – 0.1 | 0.1 – 0.2 | 0.2 – 0.3 |
| 30/s | 8.12 ± 2.38 | 19.2 ± 3.60 | 29.46 ± 4.28 |
| 60/s | 10.86 ± 3.74 | 28.0 ± 6.46 | 41.05 ± 6.14 |
| 90/s | 16.08 ± 5.25 | 35.6 ± 7.74 | 60.73 ± 7.50 |
| Mean | 11.68 ± 3.79 | 27.6 ± 5.93 | 43.75 ± 5.97 |